\documentclass[twocolumn,showpacs,showkeys,amsmath,amssymb,superscriptaddress,floatfix,nofootinbib]{revtex4}

\usepackage{graphicx}
\usepackage{bm} 
\usepackage{subfigure}
\def\vec#1{\mathchoice{\mbox{\boldmath$\displaystyle#1$}}
{\mbox{\boldmath$\textstyle#1$}}
{\mbox{\boldmath$\scriptstyle#1$}}
{\mbox{\boldmath$\scriptscriptstyle#1$}}}
\makeatletter
\newcommand\erfc{\mathop{\operator@font erfc}\nolimits}
\def\slashchar#1{\setbox0=\hbox{$#1$}
   \dimen0=\wd0 \setbox1=\hbox{/} \dimen1=\wd1
   \ifdim\dimen0>\dimen1 \rlap{\hbox to \dimen0{\hfil/\hfil}} #1
   \else  \rlap{\hbox to \dimen1{\hfil$#1$\hfil}} / \fi}

\begin{document}
 
\title{
General formulation of transverse hydrodynamics
\footnote{Supported in part by the Polish Ministry of Science and Higher Education, grant  N202 034 32/0918.}}

\author{Radoslaw Ryblewski} 
%\email{}
\affiliation{Institute of Physics, Jan Kochanowski University, PL-25406~Kielce, Poland} 

\author{Wojciech Florkowski} 
%\email{Wojciech.Florkowski@ifj.edu.pl}
\affiliation{The H. Niewodnicza\'nski Institute of Nuclear Physics, Polish Academy of Sciences, PL-31342 Krak\'ow, Poland}
\affiliation{Institute of Physics, Jan Kochanowski University, PL-25406~Kielce, Poland} 

\date{March 11, 2008}

\begin{abstract}
General formulation of hydrodynamics describing transversally thermalized matter created at the early stages of ultra-relativistic heavy-ion collisions is presented. Similarities and differences with the standard three-dimensionally thermalized relativistic hydrodynamics are discussed. The role of the conservation laws as well as the thermodynamic consistency of two-dimensional thermodynamic variables characterizing transversally thermalized matter is emphasized. 
\end{abstract}



\maketitle 

%%%%%%%%%%%%%%%%%%%%%%%%%%%%%%%%%%%%%%%%%%%%%%%%%%%%%%%%%%%%%%%%%%%%%%%%%%%%%%%%%%%%%%%%%%%%%%%%%%%%
\section{Introduction}
\label{sect:intro}
%%%%%%%%%%%%%%%%%%%%%%%%%%%%%%%%%%%%%%%%%%%%%%%%%%%%%%%%%%%%%%%%%%%%%%%%%%%%%%%%%%%%%%%%%%%%%%%%%%%%%

We present a general formulation of the hydrodynamical model describing transversally thermalized matter possibly created at the early stages of ultra-relativistic heavy-ion collisions. The concept of purely transversally thermalized matter obeying hydrodynamic evolution was initially formulated by Heinz and Wong \cite{Heinz:2002rs,Heinz:2002xf}. A new and different implementation of this idea has been recently given in Refs. \cite{Bialas:2007gn,Chojnacki:2007fi}, where, in contrast to the original formulation, it has been shown that this concept can be consistent with the RHIC data describing transverse-momentum spectra of pions and their elliptic flow coefficient $v_2$.

In this paper we extend the formal results of Refs. \cite{Bialas:2007gn,Chojnacki:2007fi} and introduce a general formalism for treating the transversally thermalized matter. Similarly to the standard relativistic hydrodynamics \footnote{By standard hydrodynamics we always mean the three-dimensionally thermalized hydrodynamics of perfect fluid.}, our formalism is based on a specific form of the conserved energy-momentum tensor. We do not restrict our consideration to the case of massless particles and the classical (Boltzmann) statistics, as it has been assumed for simplicity in \cite{Bialas:2007gn,Chojnacki:2007fi}. With the given form of the energy-momentum tensor we derive the hydrodynamic equations using the thermodynamic identities only. We also show how the discussed structure of the energy-momentum tensor and other currents follow from the explicitly covariant form of the underlying phase-space distribution function. 

Similarly to Refs. \cite{Bialas:2007gn,Chojnacki:2007fi} we view our system as a superposition of non interacting (transverse) clusters. The clusters are formed by particles moving with the same values of the rapidity. The correlation between the rapidity $y$ and the spacetime rapidity $\eta$ of the particles is built in the model by the condition
\begin{equation}
y = \eta.
\label{yeqeta}
\end{equation}

The clusters form two-dimensional (2D) objects, whose thermodynamic properties are described by appropriate 2D thermodynamic variables. One of the advantages of our approach, in comparison with the original formulation by Heinz and Wong, is that our approach is thermodynamically consistent. The 2D thermodynamic variables are derived from the formula for the potential $\Omega$ (valid for massive bosons or fermions) and satisfy standard thermodynamic identities.  

The use of the thermodynamic identities together with the conservation laws for the energy-momentum tensor leads to the entropy conservation. This behaviour is expected since the longitudinal free-streaming does not produce the entropy and the transverse expansion is entropy conserving. As a consequence, our system may be viewed as a superposition of 2D perfect fluids. 

Clearly, the microscopic collisions between the particles lead to the three-dimensional (3D) isotropization of the system, hence the proposed picture of purely transverse hydrodynamics may be valid only at the very early stages of the evolution of matter. Recently,  the possible 2D $\to$ 3D transition has been addressed in the framework of the dissipative hydrodynamics in Ref. \cite{Bozek:2007di} where the initial  asymmetric pressure relaxes towards the equilibrium isotropic form, see also Ref. \cite{Bozek:2007qt}. 

Below we discuss how our formalism may be supplemented with the 2D $\to$ 3D transition, e.g.,  by applying the Landau matching conditions between the purely transverse hydrodynamic expansion and standard 3D hydrodynamics. At the transition point the Landau conditions require local energy-momentum conservation and entropy production, which resembles the situation known from the dissipative hydrodynamics \cite{Bozek:2007di}. 

We expect that the early dynamics in relativistic heavy-ion collisions is dominated by gluons, whose number is not necessarily conserved, hence we restrict the formulation of the final equations to the case of zero chemical potential. 

Our main results are presented in Sect. \ref{sect:hydroequations2D}. The discussion of the two-dimensional thermodynamics is presented in Sect. \ref{sect:2Dthermo}. The structure of the phase-space distribution function and its moments are presented in Sect. \ref{sect:psdf}.

Below we use the following definitions for rapidity and spacetime rapidity,
\begin{eqnarray}
y = \frac{1}{2} \ln \frac{E+p_\parallel}{E-p_\parallel}, \quad
\eta = \frac{1}{2} \ln \frac{t+z}{t-z}, \label{yandeta} 
\end{eqnarray}
which come from the standard parameterization of the four-momentum and spacetime coordinate of a particle,
\begin{eqnarray}
p^\mu &=& \left(E, {\vec p}_\perp, p_\parallel \right) =
\left(m_\perp \cosh y, {\vec p}_\perp, m_\perp \sinh y \right), \nonumber \\
x^\mu &=& \left( t, {\vec x}_\perp, z \right) =
\left(\tau \cosh \eta, {\vec x}_\perp, \tau \sinh \eta \right). \label{pandx}
\end{eqnarray} 
In Eq. (\ref{pandx}) the quantity $m_\perp$ is the transverse mass
\begin{equation}
m_\perp = \sqrt{m^2 + p_x^2 + p_y^2},
\label{energy}
\end{equation}
and $\tau$ is the proper time
\begin{equation}
\tau = \sqrt{t^2 - z^2}.
\label{tau}
\end{equation} 
Throughout the paper we use the natural units where $c=1$ and $\hbar=1$.

%%%%%%%%%%%%%%%%%%%%%%%%%%%%%%%%%%%%%%%%%%%%%%%%%%%%%%%%%%%%%%%%%%%%%%%%%%%%%%%%%%%%%%%%%%%%%%%%%%%%%
\section{Hydrodynamical equations for transversally thermalized matter}
\label{sect:hydroequations2D}
%%%%%%%%%%%%%%%%%%%%%%%%%%%%%%%%%%%%%%%%%%%%%%%%%%%%%%%%%%%%%%%%%%%%%%%%%%%%%%%%%%%%%%%%%%%%%%%%%%%%%

%%%%%%%%%%%%%%%%%%%%%%%%%%%%%%%%%%%%%%%%%%%%%%%%%%%%%%%%%%%%%%%%%%%%%%%%%%%%%%%%%%%%%%%%%%%%%%%%%%%%%
\subsection{Energy-momentum conservation laws}
\label{sect:emcl}
%%%%%%%%%%%%%%%%%%%%%%%%%%%%%%%%%%%%%%%%%%%%%%%%%%%%%%%%%%%%%%%%%%%%%%%%%%%%%%%%%%%%%%%%%%%%%%%%%%%%%
Our main result is that the hydrodynamic equations for transversally thermalized matter follow from the energy-momentum conservation laws
\begin{equation}
\partial_\mu T^{\mu \nu}=0,
\label{emcl}
\end{equation}
with the energy-momentum tensor of the form
\begin{equation}
T^{\mu \nu} = \frac{n_0}{\tau} \left[
\left(\varepsilon _2 + P_2\right) U^{\mu}U^{\nu} 
- P_2 \,\,\left( g^{\mu\nu} + V^{\mu}V^{\nu} \right)\,\, \right].
\label{tensorT1}
\end{equation}
Here $n_0$ is the density of transverse clusters in rapidity and $\tau$ is the proper time defined by Eq. (\ref{tau}).

The quantities $\varepsilon_2$ and $P_2$ are the 2D energy density and 2D pressure. They are scalar quantities determined in the local rest-frame of the fluid. The subscript 2 denotes that those quantities do not refer to the 3D volume but to 2D transverse area of the cluster. Below, we also introduce the 2D entropy density $s_2$, the 2D particle density $n_2$, as well as the temperature $T$ and the chemical potential $\mu$. In Sect. \ref{sect:2Dthermo} we show in more detail how these quantities are introduced.  The important point is that all 2D thermodynamic variables are defined in the consistent way and satisfy well-known thermodynamic identities. For example, for vanishing chemical potential we have
\begin{equation}
\varepsilon_2 + P_2 = T s_2
\label{thermid1}
\end{equation}
and
\begin{equation}
d\varepsilon_2 = T ds_2, \quad dP_2 = s_2 dT.
\label{thermid2}
\end{equation}
In Eq. (\ref{tensorT1}) the four vectors $U^\mu$ and $V^\mu$ are defined by the equations:
\begin{equation}
U^{\mu} = ( u_0 \cosh\eta,u_x,u_y, u_0 \sinh\eta),
\label{U}
\end{equation}
\begin{equation}
V^{\mu} = (\sinh\eta,0,0,\cosh\eta),
\label{V}
\end{equation}
and $u^0, u_x, u_y$ are the components of the four vector
\begin{equation}
u^\mu = \left(u^0, {\vec u}_\perp, 0 \right) = \left(u^0, u_x, u_y, 0 \right), 
\label{smallu}
\end{equation}
which describes the four-velocity of the fluid element in the rest frame of the cluster to which this element belongs. The four-velocity $u^\mu$ is normalized to unity
\begin{equation}
 u^\mu u_\mu = u_0^2 - u_x^2 - u_y^2 = 1.
\label{normsmallu}
\end{equation}
The four-vector $U^\mu$ describes the four-velocity of the fluid element and plays a standard role of the hydrodynamic flow. It may be obtained from $u^\mu$ by the Lorentz boost along the $z$ axis with rapidity $\eta$ (in other words, $U^\mu$ combines the motion of the fluid element in a cluster with the motion of the cluster). The appearance of the four-vector $V^\mu$  is a new characteristic feature of transverse hydrodynamics related with the special role of the longitudinal direction -- in the rest frame of the cluster we have $V^\mu = (0,0,0,1)$.  The four vectors $U^\mu$ and $V^\mu$ satisfy the following normalization conditions:
\begin{equation}
U^\mu U_\mu = 1, \quad V^\mu V_\mu = -1, \quad U^\mu V_\mu = 0.
\label{normort}
\end{equation}
The presence of the term $V^\mu V^\nu$ in (\ref{tensorT1}) is responsible for vanishing of the longitudinal pressure. In the local rest-frame of the fluid element, where we have $U^\mu = (1,0,0,0)$ and $V^\mu = (0,0,0,1)$ one finds
\begin{equation}
T^{\mu \nu} = \frac{n_0}{\tau} \left(
\begin{array}{cccc}
\varepsilon _2 & 0 & 0 & 0 \\
0 & P_2 & 0 & 0 \\
0 & 0 & P_2 & 0 \\
0 & 0 & 0 & 0
\end{array} \right).
\end{equation}
The fact that $T^{33}=0$ indicates that there is no interaction between the clusters. Certainly, such situation cannot last long and we expect that a transition from transverse to standard 3D hydrodynamics must take place within at most few fermis after the collision.

%%%%%%%%%%%%%%%%%%%%%%%%%%%%%%%%%%%%%%%%%%%%%%%%%%%%%%%%%%%%%%%%%%%%%%%%%%%%%%%%%%%%%%%%%%%%%%%%%%%%%
\subsection{Entropy conservation}
\label{sect:entcl}
%%%%%%%%%%%%%%%%%%%%%%%%%%%%%%%%%%%%%%%%%%%%%%%%%%%%%%%%%%%%%%%%%%%%%%%%%%%%%%%%%%%%%%%%%%%%%%%%%%%%%

By straightforward calculation, we can show that the energy-momentum conservation laws (\ref{emcl}) lead to the entropy conservation, namely, we find
\begin{equation}
U_\nu \partial_\mu T^{\mu \nu} = T \partial_\mu S^\mu =0,
\label{entcl}
\end{equation}
where the entropy current is defined by the expression
\begin{equation}
S^\mu = \frac{n_0}{\tau} s_2 U^\mu.
\label{hydroS} 
\end{equation}
In Sect. \ref{sect:psdf} we show that both (\ref{tensorT1}) and (\ref{hydroS}) follow from the same phase-space distribution function, which realizes our main assumption of transverse thermalization and longitudinal free-streaming. Thus, the definition of the entropy current is consistent with the definition of the energy-momentum tensor (\ref{tensorT1}).

In the derivation of Eq. (\ref{entcl})  we used the relations
\begin{eqnarray}
\tau \, U_{\nu} V^{\mu}\partial _{\mu}V^{\nu} &=& u_0, \label{z1} \\
U^{\mu} \partial_{\mu} \tau &=& u_0, \label{z2}
\end{eqnarray}
which follow directly from the definitions (\ref{tau}), (\ref{U}) and (\ref{V}). It is worth emphasizing that Eq. (\ref{entcl}) has the structure known already from the standard relativistic hydrodynamics. In both cases the entropy conservation (the adiabaticity of the flow) is a direct consequence of the energy-momentum conservation. The situation is different in the non-relativistic hydrodynamics where the entropy conservation typically appears as an additional condition supplementing the standard Euler equation.
 
%%%%%%%%%%%%%%%%%%%%%%%%%%%%%%%%%%%%%%%%%%%%%%%%%%%%%%%%%%%%%%%%%%%%%%%%%%%%%%%%%%%%%%%%%%%%%%%%%%%%%
\subsection{Euler equations}
\label{sect:eulereq}
%%%%%%%%%%%%%%%%%%%%%%%%%%%%%%%%%%%%%%%%%%%%%%%%%%%%%%%%%%%%%%%%%%%%%%%%%%%%%%%%%%%%%%%%%%%%%%%%%%%%%

The use of the entropy conservation (\ref{entcl}) in (\ref{emcl}) leads to the equation which may be treated as the analog of the Euler equation in our case,
\begin{eqnarray}
U^\mu \partial _\mu (T U^\nu) &=& \partial ^\nu T  + V^\nu V^\mu \partial_\mu T. 
\label{hydro2D}
\end{eqnarray}
By performing projections of Eq. (\ref{hydro2D}) on the four-vectors $U_\nu$ and $V_\nu$, which yield identically zero, we verify that only two out of four equations in (\ref{hydro2D}) are independent. In this calculation we use the relations
\begin{eqnarray}
V^\mu \partial_\mu {\tau}  &=& 0, \label{z3} \\
\partial_\mu V^\mu &=& 0, \label{z4} \\
{\tau} \, V^\mu \partial_\mu V^\nu   &=& \partial^\nu {\tau}, \label{z5}
\end{eqnarray}
which, similarly as Eqs. (\ref{z1}) and (\ref{z2}), follow from the definitions (\ref{tau}), (\ref{U}) and (\ref{V}). 

From the physical point of view the fact that (\ref{hydro2D}) contains only two independent equations is expected. Since the longitudinal motion is fixed, the Euler equation may specify only the dynamics of the two transverse components of the velocity, i.e., it determines only the transverse expansion of matter in a cluster.

Altogether we have at our disposal three equations, one in (\ref{entcl}) and two in (\ref{hydro2D}), for four unknown functions: $s_2$, $T$, $u_x$ and $u_y$. To close this system we have to add the equation of state, which specifies the temperature dependence of the entropy. This situation reminds the case of the standard relativistic hydrodynamics where we have four equations for five unknown functions and also the equation of state must be added to close the system of equations \footnote{This is so in the case of vanishing baryon potential. For non-zero baryon potential we have one more thermodynamic variable (baryon number density) and one more equation (the baryon number conservation law).}.

%%%%%%%%%%%%%%%%%%%%%%%%%%%%%%%%%%%%%%%%%%%%%%%%%%%%%%%%%%%%%%%%%%%%%%%%%%%%%%%%%%%%%%%%%%%%%%%%%%%%%
\subsection{Cylindrical coordinates}
\label{sect:eulereq}
%%%%%%%%%%%%%%%%%%%%%%%%%%%%%%%%%%%%%%%%%%%%%%%%%%%%%%%%%%%%%%%%%%%%%%%%%%%%%%%%%%%%%%%%%%%%%%%%%%%%%

The three independent equations of transverse hydrodynamics may be conveniently written in the cylindrical coordinates. In this case we introduce the distance $r$ and the azimuthal angle $\phi$ defined by the equations,
\begin{equation}
r\!\!=\!\!\sqrt{r_x^2+r_y^2}, \quad \phi=\hbox{tan}^{-1} (r_y/r_x),
\label{cylinder}
\end{equation}
where ${\vec x}_\perp = (r_x,r_y)$. In the similar way we parameterize the fluid velocity, 
\begin{eqnarray}
v_x &=& v \cos(\alpha+\phi), \quad v_y = v \sin(\alpha+\phi), \label{vxy} \\
u_x &=& u_0 v_x = u_\perp \cos(\alpha+\phi), \label{ux} \\
u_y &=& u_0 v_y = u_\perp \sin(\alpha+\phi). \label{uy}
\label{vxvy}
\end{eqnarray}
Here $v$ is the transverse flow, $u_0 =\left(1-v^2\right)^{-\frac{1}{2}}$, $u_\perp = u_0 v$, and the dynamical angle $\alpha$ describes deviations of the flow direction from the radial direction.

With the help of such variables the hydrodynamic equations may be written explicitly in the following form
\begin{eqnarray}
&& \frac{\partial }{\partial \tau} \left( r s_2 u_0 \right) +\frac{\partial }{\partial r}
\left( r s_2 u_0 v\cos \alpha \right) + \frac{\partial }{\partial
\phi }\left( s_2 u_0 v\sin \alpha \right)  =0, \nonumber \\
&& \frac{\partial }{\partial \tau}\left( rTu_0 v\right) +
r\cos \alpha \frac{\partial }{\partial r}\left( Tu_0 \right) 
+\sin \alpha \frac{\partial }{\partial \phi }\left( Tu_0 \right)  =0, \nonumber \\
&& Tu_0 ^{2}v\left( \frac{d\alpha }{d \tau}+\frac{v\sin \alpha }{r}\right) -\sin
\alpha \frac{\partial T}{\partial r}
+\frac{\cos \alpha }{r}\frac{\partial T}{\partial \phi } =0. 
\label{wfdyr3}
\end{eqnarray}
In (\ref{wfdyr3}) the derivative $d/d\tau$ denotes the total derivative with respect to time
\begin{eqnarray}
\frac{d}{d\tau}=\frac{\partial}{\partial \tau} + v \cos \alpha \frac{\partial}{\partial r} + \frac{v \sin \alpha }{r} \frac{\partial}{\partial \phi} .
\label{totaltd}
\end{eqnarray}  
We note that Eqs. (\ref{wfdyr3}) have exactly the same structure as Eqs. (20) in Ref. \cite{Chojnacki:2007fi}. The important point is, however, that they were derived here directly from the energy-momentum tensor (\ref{emcl}) with the help of the thermodynamic identities only. No specific dependence of the thermodynamic variables on the temperature was used.  

%%%%%%%%%%%%%%%%%%%%%%%%%%%%%%%%%%%%%%%%%%%%%%%%%%%%%%%%%%%%%%%%%%%%%%%%%%%%%%%%%%%%%%%%%%%%%%%%%%%%%
\subsection{Breaking of boost-invariance}
\label{sect:eulereq}
%%%%%%%%%%%%%%%%%%%%%%%%%%%%%%%%%%%%%%%%%%%%%%%%%%%%%%%%%%%%%%%%%%%%%%%%%%%%%%%%%%%%%%%%%%%%%%%%%%%%%

So far we treated the cluster density $n_0$ as a constant parameter. One may check, however, that $n_0$ may depend on the spacetime rapidity $\eta$. To see this point, we write the the energy-momentum tensor as the product 
\begin{equation}
T^{\mu \nu} = n_0(\eta) T^{\mu \nu}_{(1)}.
\end{equation}
The energy-momentum tensor $T^{\mu \nu}_{(1)}$ agrees with our previous form for the case $n_0 = 1$ and is conserved, 
\begin{equation}
\partial_\mu T^{\mu \nu}_{(1)} = 0.
\end{equation}
It turns out that the conservation of the tensor $T^{\mu \nu}_{(1)}$ implies the conservation of the tensor $T^{\mu \nu}$. To prove this point we consider the expression $\partial_\mu T^{\mu \nu}$. Since the tensor $T^{\mu \nu}_{(1)}$ is conserved, we find
\begin{equation}
\partial_\mu T^{\mu \nu} = \partial_\mu \left[ n_0(\eta) T^{\mu \nu}_{(1)} \right] =
T^{\mu \nu}_{(1)} \partial_\mu n_0(\eta).
\end{equation}
With the explicit form of  $T^{\mu \nu}_{(1)}$ we get
\begin{eqnarray}
\partial_\mu T^{\mu \nu} &=&
\frac{1}{\tau} 
\left(\varepsilon _2 + P_2\right) U^{\mu}U^{\nu} \partial_\mu n_0(\eta)
\nonumber \\
&& -\frac{1}{\tau} P_2 \,\,\left( g^{\mu\nu} + V^{\mu}V^{\nu} \right)
\partial_\mu n_0(\eta).
\label{dT1}
\end{eqnarray}
Since the operator $U^\mu \partial_\mu$ does not contain the derivative with respect to $\eta$, the non-zero contributions on the right-hand side of Eq. (\ref{dT1}) may come only from the terms in the second line with $\nu = 0$ or $\nu = 3$. The explicit calculation shows, however, that these terms yield zero, hence the tensor $T^{\mu \nu}$ is conserved.
 
The fact that $n_0$ may depend on $\eta$ indicates that our formalism is not necessarily boost-invariant. Actually, a more detailed analysis shows that our three hydrodynamic equations describe the expansion of matter in a single cluster and different initial conditions may be applied for different values of $\eta$. Such situation is of course a direct consequence of our main assumption
that clusters are independent. 

%%%%%%%%%%%%%%%%%%%%%%%%%%%%%%%%%%%%%%%%%%%%%%%%%%%%%%%%%%%%%%%%%%%%%%%%%%%%%%%%%%%%%%%%%%%%%%%%%%%%%
\subsection{Landau matching conditions at 2D $\to$ 3D transition}
\label{sect:2D3d}
%%%%%%%%%%%%%%%%%%%%%%%%%%%%%%%%%%%%%%%%%%%%%%%%%%%%%%%%%%%%%%%%%%%%%%%%%%%%%%%%%%%%%%%%%%%%%%%%%%%%%

The 2D $\to$ 3D transition may be described by assuming the Landau matching conditions at the transition point
\begin{equation}
T^{\mu \nu} U_\nu = T_{3D}^{\mu \nu} U_\nu,
\label{LMC}
\end{equation}
where $T_{3D}^{\mu \nu}$ is the standard energy-momentum tensor of relativistic hydrodynamics of perfect fluid

\begin{equation}
T_{3D}^{\mu \nu} = \left(\varepsilon + P\right) U^\mu U^\nu - P g^{\mu \nu}.
\label{standardTmunu}
\end{equation}
Substitution of Eqs. (\ref{tensorT1}) and (\ref{standardTmunu}) into Eq. (\ref{LMC}) gives
\begin{equation}
\frac{n_0}{\tau} \varepsilon_2 U^\mu  = \varepsilon U^\mu.
\label{LMC1}
\end{equation}
This condition specifies nothing else but the local conservation of energy and momentum at the transition point. In addition, Eq. (\ref{LMC1}) must be supplemented by the condition that the entropy increases
\begin{equation}
\frac{n_0}{\tau} s_2 < s.
\end{equation}
Here $s$ is the 3D entropy density corresponding to the 3D energy density $\varepsilon$.

We note that a similar procedure of Landau matching has been recently used in Ref. \cite{Broniowski:2008vp}, where the 3D free-streaming in the early stage is followed by the
3D boost-invariant hydrodynamics.

%%%%%%%%%%%%%%%%%%%%%%%%%%%%%%%%%%%%%%%%%%%%%%%%%%%%%%%%%%%%%%%%%%%%%%%%%%%%%%%%%%%%%%%%%%%%%%%%%%%%%
%%%%%%%%%%%%%%%%%%%%%%%%%%%%%%%%%%%%%%%%%%%%%%%%%%%%%%%%%%%%%%%%%%%%%%%%%%%%%%%%%%%%%%%%%%%%%%%%%%%%%
\section{Thermodynamics of two-dimensional systems}
\label{sect:2Dthermo}
%%%%%%%%%%%%%%%%%%%%%%%%%%%%%%%%%%%%%%%%%%%%%%%%%%%%%%%%%%%%%%%%%%%%%%%%%%%%%%%%%%%%%%%%%%%%%%%%%%%%%
%%%%%%%%%%%%%%%%%%%%%%%%%%%%%%%%%%%%%%%%%%%%%%%%%%%%%%%%%%%%%%%%%%%%%%%%%%%%%%%%%%%%%%%%%%%%%%%%%%%%%

The starting point of our thermodynamic considerations is the formula for the potential $\Omega$ of non-interacting bosons (upper signs) or fermions (lower signs). In the case of two-dimensional systems it has the following form 
\begin{equation}
\Omega(T,V_2,\mu) = \pm \nu _g T V_2 \int \frac{d^2 p_{\perp} }{(2\pi)^2}\,
\ln \left(1 \mp e^{(\mu-m_\perp)/T} \right).
\label{Omega1}
\end{equation}
The number of particles is not conserved in our approach and we assume $\mu =0$ (however, this can be done only in the end of the thermodynamic calculations). Since the motion of particles is confined to a plane, their energy is equal to their transverse mass defined by Eq. (\ref{energy}).

The quantity $V_2$ is the area of the plane where the particles move and $\nu_g$ denotes the internal degrees of freedom. For gluon dominated system $\nu_g = 16$. Eq. (\ref{Omega1}) allows us to define the particle density $n_2 = N_2/V_2$, pressure $P_2$,  entropy density $s_2 = S_2/V_2$, and energy density $ \varepsilon_2  = E_2/V_2$. Those quantities follow from the well-known thermodynamic relations:
\begin{eqnarray}
N_2 &=& - \left( \frac{ \partial \Omega}{ \partial \mu} \right) _{V_2 , T}, 
\label{N2} \\
P_2 &=& - \left( \frac{ \partial \Omega}{ \partial V_2} \right) _{T, \mu} = - \frac{\Omega}{V_2},    
\label{P2} \\
S_2 &=& - \left( \frac{ \partial \Omega}{ \partial T} \right) _{\mu, V_2},
\label{S2} 
\end{eqnarray}
and
\begin{equation}
E_2 + P_2 V_2 = T S_2 + \mu \, N_2.
\label{GD1}
\end{equation}
In our calculations the last equation is most often used in the form valid for the densities, see Eq. (\ref{thermid1}). In the case $\mu = 0$  Eqs. (\ref{N2}) - (\ref{GD1}) yield:
\begin{eqnarray}
\!\!\!\!\!\!\!\!N_2 &=& \nu _g V_2 \int \frac{d^2 p_\perp}{(2\pi)^2} \,g,  \label{N21} \\
\!\!\!\!\!\!\!\!P_2 &=& \nu _g  \int \frac{d^2 p_\perp}{(2\pi)^2} \frac{p_\perp^2}{2 m_\perp} \,g, 
\label{P21} \\
\!\!\!\!\!\!\!\!S_2 &=& - \nu _g V_2 \int \frac{d^2 p_\perp}{(2\pi)^2} \left[g \ln g
\mp \left(1 \pm g \right) \ln \left(1 \pm g \right) \right] ,  \label{S21} \\
\!\!\!\!\!\!\!\!E_2 &=& \nu _g V_2 \int \frac{d^2 p_\perp}{(2\pi)^2} \,m_\perp \,g,  \label{E21}
\end{eqnarray}
where we have introduced the equilibrium distribution function
\begin{equation}
g(m_\perp) = \frac{1}{e^{m_\perp/T} \mp 1}.
\label{geq}
\end{equation}
Here again the upper (lower) signs refer to bosons (fermions).

%%%%%%%%%%%%%%%%%%%%%%%%%%%%%%%%%%%%%%%%%%%%%%%%%%%%%%%%%%%%%%%%%%%%%%%%%%%%%%%%%%%%%%%%%%%%%%%%%%%%%
\subsection{Massless fermions and bosons}
%%%%%%%%%%%%%%%%%%%%%%%%%%%%%%%%%%%%%%%%%%%%%%%%%%%%%%%%%%%%%%%%%%%%%%%%%%%%%%%%%%%%%%%%%%%%%%%%%%%%%

In the case of massles fermions or bosons, the integrals (\ref{N21}) and (\ref{E21}) are analytic and yield:
\begin{eqnarray}
n_{2} &=& \frac{\nu_g \pi T^2}{24},  \quad 
\varepsilon_{2} = \frac{3 \nu_g  \zeta(3) T^3}{4 \pi} \quad \hbox{(fermions)}
\label{fermions} 
\end{eqnarray}
and
\begin{eqnarray}
n_{2} &=& \frac{\nu_g \pi T^2}{12},  \quad 
\varepsilon_{2} = \frac{ \nu_g  \zeta(3) T^3}{\pi} \quad \hbox{(bosons)}.
\label{bosons} 
\end{eqnarray}
In both cases we also find
\begin{eqnarray}
P_2 = \frac{1}{2} \varepsilon_2,  \quad 
c_s^2 = \frac{\partial P_2}{\partial \varepsilon_2} = \frac{1}{2}. \label{pressure} 
\end{eqnarray}
The entropy density may be obtained from Eq. (\ref{thermid1}) with the appropriate substitutions. 

%%%%%%%%%%%%%%%%%%%%%%%%%%%%%%%%%%%%%%%%%%%%%%%%%%%%%%%%%%%%%%%%%%%%%%%%%%%%%%%%%%%%%%%%%%%%%%%%%%%%%
\subsection{Classical limit}
%%%%%%%%%%%%%%%%%%%%%%%%%%%%%%%%%%%%%%%%%%%%%%%%%%%%%%%%%%%%%%%%%%%%%%%%%%%%%%%%%%%%%%%%%%%%%%%%%%%%%

In the case of the classical (Boltzmann) statistics the analytic formulas for the densities of thermodynamic quantities may be easily obtained also for finite masses,
\begin{eqnarray}
n_2 &=& \frac{ \nu _g T}{2\pi} (m+T) e ^{ -m/T },  \label{en2} \\
P_2 &=& \frac{ \nu _g T^2}{2\pi} (m+T) e ^{ -m/T }, \label{pe2} \\
s_2 &=& \frac{ \nu _g}{2\pi} [m^2+3mT+3T^2] e ^{ -m/T}, \label{es2} \\
\varepsilon_2 &=& \frac{ \nu _g T }{2\pi} [T^2+(m+T)^2] e ^{ -m/T }.  \label{eps2}
\end{eqnarray}
In the limit $m \to 0$ these results agree with the expression first given in Ref. 
\cite{Bialas:2007gn}. we note that in the case of the Boltzmann statistics, the equation of state is of the well-known form,
$P_2 V_2 = N_2 T$. With the help of the thermodynamic identities $d\varepsilon = T ds_2$ and $dP_2 = s_2 dT$ we find the formula for the sound velocity,
\begin{eqnarray}
c_s^2 &=& \frac{ \partial P_2}{ \partial \varepsilon _2} = \frac{ s_2 dT}{T ds_2} =
\frac{d\ln T}{ d\ln s_2} \nonumber \\
&=&  \frac{T(m^2+3 m T + 3 T^2)}{m^3 + 3 m^2 T + 6 m T^2 + 6 T^3}.
\label{cs2-1}
\end{eqnarray}
In the limiting cases one obtains the expected results,
\begin{eqnarray}
\lim\limits_{m \to 0}  c_s^2  &=& \frac{1}{2}, \quad \lim\limits_{T \to \infty}  
c_s^2  = \frac{1}{2}.
\end{eqnarray}

%%%%%%%%%%%%%%%%%%%%%%%%%%%%%%%%%%%%%%%%%%%%%%%%%%%%%%%%%%%%%%%%%%%%%%%%%%%%%%%%%%%%%%%%%%%%%%%%%%%%%
%%%%%%%%%%%%%%%%%%%%%%%%%%%%%%%%%%%%%%%%%%%%%%%%%%%%%%%%%%%%%%%%%%%%%%%%%%%%%%%%%%%%%%%%%%%%%%%%%%%%%
\section{Lorentz structure of the phase-space distribution function and its moments}
\label{sect:psdf}
%%%%%%%%%%%%%%%%%%%%%%%%%%%%%%%%%%%%%%%%%%%%%%%%%%%%%%%%%%%%%%%%%%%%%%%%%%%%%%%%%%%%%%%%%%%%%%%%%%%%%
%%%%%%%%%%%%%%%%%%%%%%%%%%%%%%%%%%%%%%%%%%%%%%%%%%%%%%%%%%%%%%%%%%%%%%%%%%%%%%%%%%%%%%%%%%%%%%%%%%%%%

In this Section we analyze in more detail the Lorentz structure of the phase-space distribution function which leads to the form of the energy-momentum tensor discussed in Sect. II.

%%%%%%%%%%%%%%%%%%%%%%%%%%%%%%%%%%%%%%%%%%%%%%%%%%%%%%%%%%%%%%%%%%%%%%%%%%%%%%%%%%%%%%%%%%%%%%%%%%%%%
%%%%%%%%%%%%%%%%%%%%%%%%%%%%%%%%%%%%%%%%%%%%%%%%%%%%%%%%%%%%%%%%%%%%%%%%%%%%%%%%%%%%%%%%%%%%%%%%%%%%%
\subsection{Distribution function}
\label{sect:distribution}
%%%%%%%%%%%%%%%%%%%%%%%%%%%%%%%%%%%%%%%%%%%%%%%%%%%%%%%%%%%%%%%%%%%%%%%%%%%%%%%%%%%%%%%%%%%%%%%%%%%%%
%%%%%%%%%%%%%%%%%%%%%%%%%%%%%%%%%%%%%%%%%%%%%%%%%%%%%%%%%%%%%%%%%%%%%%%%%%%%%%%%%%%%%%%%%%%%%%%%%%%%%

According to the original formulation of the model in Ref. \cite{Bialas:2007gn}, we assume that the phase-space distribution function $F$ may be factorized into the longitudinal and transverse part
\begin{equation}
F  = f_{\parallel} \, g.
\label{Fxp}
\end{equation}
In the global equilibrium, the function $g$ agrees with the equilibrium distribution (\ref{geq}). In the local equilibrium, assumed in the case of the transverse hydrodynamic expansion, it has the form 
\begin{equation}
g\left(p^\mu U_\mu\right) = \frac{1}{e^{p^\mu U_\mu/T} \mp 1},
\label{geqh}
\end{equation}
where 
\begin{equation}
p^\mu U_\mu = m_\perp u_0 \cosh(y-\eta) - {\vec p}_\perp \cdot {\vec u}_\perp.
\label{pdotU}
\end{equation}
The longitudinal part of the distribution function is given by the expression which implements the condition (\ref{yeqeta}), 
\begin{eqnarray}
f_{\parallel} =  n_0 \frac{ \delta ( y - \eta )}{m_{\perp } \tau} =
\frac{n_0}{\tau } \delta \left( p^{\mu} V_{\mu} \right).
\label{fpar-1}
\end{eqnarray}
%
%\ bigskip
%
Combining Eqs. (\ref{Fxp}), (\ref{geqh}) and (\ref{fpar-1}) we obtain the form of the phase-space distribution function which explicitly emphasizes its scalar character, 
\begin{equation}
F  = \frac{n_0}{\tau } \delta \left( p \cdot V \right) g\left( p \cdot U \right).
\label{Fxp1}
\end{equation}

%%%%%%%%%%%%%%%%%%%%%%%%%%%%%%%%%%%%%%%%%%%%%%%%%%%%%%%%%%%%%%%%%%%%%%%%%%%%%%%%%%%%%%%%%%%%%%%%%%%%%
%%%%%%%%%%%%%%%%%%%%%%%%%%%%%%%%%%%%%%%%%%%%%%%%%%%%%%%%%%%%%%%%%%%%%%%%%%%%%%%%%%%%%%%%%%%%%%%%%%%%%
\subsection{Particle current}
\label{sect:particlecurrent}
%%%%%%%%%%%%%%%%%%%%%%%%%%%%%%%%%%%%%%%%%%%%%%%%%%%%%%%%%%%%%%%%%%%%%%%%%%%%%%%%%%%%%%%%%%%%%%%%%%%%%
%%%%%%%%%%%%%%%%%%%%%%%%%%%%%%%%%%%%%%%%%%%%%%%%%%%%%%%%%%%%%%%%%%%%%%%%%%%%%%%%%%%%%%%%%%%%%%%%%%%%%

The particle current $N^\mu$ is defined as the first moment of the distribution function,
\begin{equation}
N^{\mu} = \frac{n_0 \nu _g }{ (2\pi)^2 \tau}   \int \frac{d^3p}{p^0} p^{\mu} 
\delta (p \cdot V ) g(p \cdot U).
\end{equation}  
The Lorentz structure of the distribution function implies that $N^\mu$ may be written as the combination of the four-vectors $V^\mu$ and $U^\mu$, namely
\begin{equation}
N^{\mu} = a \, V^{\mu} + b \, U^{\mu}.
\end{equation}  
The coefficients $a$ and $b$ are obtained by the projections, 
\begin{equation}
V_{\mu} N^\mu = -a, \quad  U_{\mu} N^\mu = b,
\end{equation}  
and, as scalar quantities, they may be calculated in the local rest frame of the fluid element where \mbox{$U^\mu = (1,0,0,0)$} and $V^\mu = (0,0,0,1)$. The explicit calculations give
\begin{equation}
a = - \frac{n_0 \nu _g}{\tau (2\pi)^2}   \int \frac{d^3p}{E} \, p_{\parallel} \, \delta (p_{\parallel}) g(E) = 0,
\end{equation}  
\begin{equation}
b = \frac{n_0 \nu _g}{\tau (2\pi)^2}   \int \frac{d^3p}{E} \,E \, \delta (p_{\parallel}) g(E) = \frac{n_0}{\tau} n_2.
\end{equation}  
Here we used the definition of the two-dimensional particle density following from Eq. (\ref{N21}). Combining all the results of this Section we find the general structure of the particle current,
\begin{equation}
N^\mu = \frac{n_0}{\tau} n_2 U^\mu
\end{equation}

%%%%%%%%%%%%%%%%%%%%%%%%%%%%%%%%%%%%%%%%%%%%%%%%%%%%%%%%%%%%%%%%%%%%%%%%%%%%%%%%%%%%%%%%%%%%%%%%%%%%%
%%%%%%%%%%%%%%%%%%%%%%%%%%%%%%%%%%%%%%%%%%%%%%%%%%%%%%%%%%%%%%%%%%%%%%%%%%%%%%%%%%%%%%%%%%%%%%%%%%%%%
\subsection{Energy-momentum tensor}
\label{sect:emt}
%%%%%%%%%%%%%%%%%%%%%%%%%%%%%%%%%%%%%%%%%%%%%%%%%%%%%%%%%%%%%%%%%%%%%%%%%%%%%%%%%%%%%%%%%%%%%%%%%%%%%
%%%%%%%%%%%%%%%%%%%%%%%%%%%%%%%%%%%%%%%%%%%%%%%%%%%%%%%%%%%%%%%%%%%%%%%%%%%%%%%%%%%%%%%%%%%%%%%%%%%%%

The energy-momentum tensor is the second moment of the distribution function. It should have the following Lorentz structure, 
\begin{equation}
 T_{\mu \nu} = a^{\, \prime} U_{\mu}U_{\nu} + b^{\, \prime} g_{\mu \nu} 
+ c^{\, \prime} V_{\mu} V_{\nu} + \frac{d^{\, \prime}}{2} (U_{\mu}V_{\nu} + U_{\nu}V_{\mu}).
\label{Tdec}
\end{equation}
Here, besides the symmetric combinations of the four-vectors $U^\mu$ and $V^\mu$, the term proportional to the metric tensor is also required. Simple algebra gives
\begin{eqnarray}
T^{\mu}_{\hspace{1.5mm} \mu} &=&  a^{\, \prime} + 4 b^{\, \prime} - c^{\, \prime},  \nonumber \\
T^{\mu \nu} U_{\mu}U_{\nu} &=& a^{\, \prime} + b^{\, \prime},  \nonumber \\
T^{\mu \nu} V_{\mu}V_{\nu} &=& - b^{\, \prime} + c^{\, \prime}, \nonumber \\
T^{\mu \nu} U_{\mu}V_{\nu} &=& T^{\mu \nu} U_{\nu}V_{\mu} = 
- \frac{d^{\, \prime}}{2}. 
\label{wsp0}
\end{eqnarray}
The coefficients in the decomposition (\ref{Tdec}) may be again calculated in the local rest frame of the fluid element. This procedure yields immediately the two constraints: $d^{\, \prime}=0$ and $c^{\, \prime}=b^{\, \prime}$.  Thus, Eqs. (\ref{wsp0}) are reduced to the form
\begin{eqnarray}
T^{\mu}_{\hspace{1.5mm} \mu} &=& a^{\, \prime} + 3 b^{\, \prime},  \nonumber \\
T^{\mu \nu} U_{\mu}U_{\nu} &=& a^{\, \prime} + b^{\, \prime},  \nonumber \\
c^{\, \prime} &=& b^{\, \prime}  \nonumber \\
d^{\, \prime} &=& 0.
\label{B-3}
\end{eqnarray}
The trace of the energy-momentum tensor equals
\begin{equation}
T^{\mu}_{\hspace{2mm}\mu} =
\frac{n_0 \nu _g}{(2\pi)^2 \tau}  \int \frac{d^3p}{p^0} m^2  
\delta (p \cdot V) g(p \cdot U),
\end{equation}
where we used $p^{\mu} p_{\mu} = m^2$. In the local rest frame we find 
\begin{eqnarray}
T^{\mu}_{\hspace{2mm} \mu} &=& \frac{n_0  \nu _g }{ (2\pi)^2 \tau }  
\int \frac{d^3p}{E}  m^2 \delta (p_{\parallel}) g(E)  \nonumber \\
&=& \frac{n_0 \nu _g }{ \tau }  \int \frac{d^2 p_{\perp}}{(2 \pi)^2} 
\frac{m^2}{m_{\perp }} g(m_{\perp}) \nonumber \\
&=& \frac{n_0}{\tau} (\varepsilon_2 - 2 P_2).
\label{B-4}
\end{eqnarray}
Here we used Eqs. (\ref{P21}) and (\ref{E21}). We note that for massless particles the trace vanishes, $T^{\mu}_{\hspace{2mm} \mu} = 0$, and we find $\varepsilon_2 = 2 P_2$. This in turn gives
$c_s^2 = 1/2$. In the next step we calculate the projection of the energy-momentum tensor on the four-velocity $U^\mu$, 
\begin{eqnarray}
T^{\mu \nu} U_{\mu}U_{\nu} &=& \frac{n_0 \nu _g}{(2\pi)^2 \tau }  
\int \frac{d^3p}{p^0} (p \cdot U)^2  \delta (p \cdot V) \, 
g(p \cdot U)  \nonumber \\
&=& \frac{n_0 \nu _g}{(2\pi)^2 \tau }  \int d^2 p_{\perp} m_{\perp }  g(m_{\perp})
= \frac{n_0}{\tau} \varepsilon_2.
\label{TUU}
\end{eqnarray}
Eqs.(\ref{B-3}) - (\ref{TUU}) lead us directly to our basic definition (\ref{tensorT1}).

%%%%%%%%%%%%%%%%%%%%%%%%%%%%%%%%%%%%%%%%%%%%%%%%%%%%%%%%%%%%%%%%%%%%%%%%%%%%%%%%%%%%%%%%%%%%%%%%%%%%%
%%%%%%%%%%%%%%%%%%%%%%%%%%%%%%%%%%%%%%%%%%%%%%%%%%%%%%%%%%%%%%%%%%%%%%%%%%%%%%%%%%%%%%%%%%%%%%%%%%%%%
\subsection{Entropy current}
\label{sect:entropy}
%%%%%%%%%%%%%%%%%%%%%%%%%%%%%%%%%%%%%%%%%%%%%%%%%%%%%%%%%%%%%%%%%%%%%%%%%%%%%%%%%%%%%%%%%%%%%%%%%%%%%
%%%%%%%%%%%%%%%%%%%%%%%%%%%%%%%%%%%%%%%%%%%%%%%%%%%%%%%%%%%%%%%%%%%%%%%%%%%%%%%%%%%%%%%%%%%%%%%%%%%%%

Finally, we consider the entropy current,
\begin{eqnarray}
S^{\mu} &=& -\frac{n_0 \nu _g}{(2\pi)^2 \tau } \\
&&  \times \int \frac{d^3p}{p^0} p^{\mu}\,\,\delta (p \cdot V)  
g(p \cdot U) [\,\, \ln[ g ( p \cdot U) ] - 1\,\, ]. \nonumber
\end{eqnarray}
The meaning of this definition is that we use the Boltzmann expression to calculate the entropy of a single cluster and then the sum over clusters is performed. Similarly to the case of the particle current we express the entropy current as a linear combination of the four-vectors $U^\mu$ and $V^\mu$,
\begin{equation}
S^{\mu} = a^{\, \prime \prime} V^{\mu} + b^{\, \prime \prime} U^{\mu}.
\label{Scom}
\end{equation}
In the analogous way to the cases studied above we find
\begin{equation}
a^{\, \prime \prime}=0, \quad  b^{\, \prime \prime} = \frac{n_0}{\tau} s_2,
\label{Swsp}
\end{equation}
hence the definition (\ref{hydroS}) is recovered. 

\section{Conclusions}

In this paper we have introduced the general formalism for hydrodynamical description of transversally thermalized matter possibly created in the early stages of ultra-relativistic heavy-ion collisions. We have showed how the hydrodynamic equations follow directly from the general form of the energy-momentum tensor. We discussed the entropy conservation law and the structure of the underlying phase-space distribution. The thermodynamics of two-dimensional systems has been analyzed and the generally covariant structure of the model has been emphasized. After first successful attempts to describe physical observables studied at RHIC \cite{Bialas:2007gn} we hope that the presented formalism will find further applications to describe early stages of relativistic heavy-ion collisions.  

Acknowledgments: We thank Andrzej Bialas and \mbox{Wojtek} Broniowski for critical and useful comments concerning the manuscript.  

%\bibliography{ref-rr}

\end{document}